
.............................................................

Special macros are necessary to print this file. Figures are not
attached. Please contact the authors if you need a hardcopy.
.............................................................

\input jnl.tex
\input reforder.tex
\input eqnorder

\def\jt{\rm J_z/t}

\def\dx2y2{\rm d_{x^2-y^2}}

\null
\vskip -.75in
\vskip .5in
{\singlespace
\smallskip
\rightline{ November, 1992}
}

\vskip 0.1in

\title BASIS SET REDUCTION APPLIED TO
       THE TWO DIMENSIONAL ${\rm t-J_z}$ MODEL

\vskip 0.2in

\author  J.~RIERA$^\ast$ and E.~DAGOTTO

\affil Department of Physics,
       Center for Materials Research and Technology and
       Supercomputer Computations Research Institute,
       Florida State University,
       Tallahassee, FL 32306

\vskip 0.1in

\abstract { A simple variation of the Lanczos method is discussed. The
new technique is based on a systematic reduction of the size of
the Hilbert space of the model under consideration. As an example,
the two dimensional
${\rm t-J_z}$ model of strongly correlated electrons is studied.
Accurate
results for the ground state energy
can be obtained on clusters of up to 50
sites, which are unreachable by conventional Lanczos approaches.
In particular, the energy of one and two holes is analyzed as a function
of ${\rm J_z/t}$. In the bulk limit, it is shown that a finite coupling
${\rm J_z/t ]_c} \sim 0.18$ is
necessary to induce ``binding'' of holes in the model. It is argued that
this result implies that the two dimensional ${\rm t-J}$ model
phase separates only for couplings larger than a $finite$ critical value.
}

\vskip 0.4truecm

\line{PACS Indices: 75.10.Jm, 75.40.Mg, 74.20.-z\hfill }

\vskip 0.4truecm

{\singlespace
\noindent
$\ast$) Permanent address:
Departamento de F\'isica, Facultad de Ciencias Exactas,
Av. Pellegrini 250, 2000 Rosario, Argentina.}

\endtitlepage

The study of
strongly correlated electrons is currently
one of the most active areas of
research in condensed matter and materials science.
Several important experimental
discoveries in the last few years have clearly shown that interesting
new physics in various compounds can appear in the regime where electrons
interact $strongly$. Typical examples are high temperature
superconductors, ${\rm C_{60}}$ fullerenes, and heavy
fermions.\refto{muller} However, in spite of this significant experimental
progress, a theoretical explanation for the behavior of electrons
in these materials
is still unknown.
Appealing ``scenarios'' abound, but their
predictions are
in general based on rough uncontrollable approximations.
After analyzing in detail the various theoretical proposals,
it becomes clear that an
accurate technique to carry out the
calculations in strong coupling is needed,
since in this regime perturbation theory is questionable. No natural
small parameter controls the comparison between
theoretical predictions and experiments.
In an attempt to reduce this theory-experiment gap,
in recent years computational studies have
become popular in the context of interacting electrons.
In principle, to analyze these models numerically
there is no need to have a small parameter. However, the
main problem in the application of
these techniques comes from ``finite size effects.''
As a typical example, let us consider
the Lanczos method\refto{review} which
has proven to be very effective in the evaluation
of static and dynamical properties of complicated models defined on
finite clusters. Extensive analysis by several groups have shown that
the qualitative predictions obtained with this approach are usually correct.
However, obtaining more quantitative results
in the bulk limit is difficult due to the rapid growth of the basis set
of realistic models with the size of the lattice.
Other methods, like
Quantum Monte Carlo, present the complication of the
``sign-problem'' that has plagued simulations of Hubbard-like models
at finite electron density and low temperatures.
Then, any progress on the implementation of
numerical methods may represent an important step forward in the
understanding of theories of interacting electrons.

The purpose of this paper is to present a simple modification of the
widely used Lanczos technique, that allows the study of large clusters for
some particular Hamiltonians.
The standard Lanczos method basically consists of an iterative
procedure to diagonalize a large
Hamiltonian matrix corresponding to electrons on a finite
cluster. While the technique is extremely accurate and flexible,
and allows the study of dynamical response functions in real time, its
main constraint comes from the exponential growth of the Hilbert space
with the number of sites.
For example, for the two dimensional ${\rm t-J}$ model,
presumed to describe in some regime high
temperature superconductors, the maximum cluster size analyzed in the
literature corresponds to one hole on a 26 sites
lattice.\refto{didier}
However, in some cases it has been observed that many coefficients $|c_n|$
in the expansion of the ground state wave function in
the ${\rm S_z}$ basis, i.e.
$| \psi_0 \rangle = \sum_{n} c_n |n \rangle$, are negligible.
Then,
it may occur that restricting the diagonalization of the problem to
a $reduced$ basis set may still provide accurate enough results. Work
along these lines
have been recently applied to the analysis of the Hubbard
model,\refto{germans} and carbon atoms and beryllium dimers.\refto{wilson}

The algorithm proposed
in this paper consists of the following steps:
first, a ``simple'' state is used to start the iterations. This state is
selected such that it already belongs to the representation of the
symmetry group corresponding to the ground state. For example, the
momentum of the state is fixed by working in an appropriate basis
of momentum eigenstates. If other symmetries are available and implemented, the
convergence rate of the technique is improved.
After the initial state is selected, the hopping term of the
Hamiltonian is applied several times in order to produce a basis
set of, e.g., a few hundred states. At this point, a Lanczos
diagonalization within that reduced space is carried out. The
wave function of the ground state of this reduced subspace is analyzed, and
basis states with $|c_n|$ smaller than a given parameter $\lambda$
are simply discarded. $\lambda$
is selected such that the number of basis states necessary to expand
the new $truncated$ ground state
is only slightly larger than the basis in the previous iteration.
This procedure is repeated several times, i.e., first,
the Hilbert space is further expanded by the application of the hopping
term to the truncated ground state of the previous iteration.
Second, a Lanczos diagonalization is performed in the new
enlarged space.\refto{trugman}
The basis set of the lowest energy state in this
enlarged space is ``pruned'' according to the criterion $|c_n| < \lambda$, and,
finally, a new iteration starts again.\refto{foot3}
As a nontrivial model to test these ideas, the ${\rm t-J_z}$ in two
dimensions was selected. It describes holes carriers moving in an
antiferromagnetic
background. This model is defined by the Hamiltonian,\refto{string,barnes}
$$
{\rm H =
{\rm J_z } \sum_{{\bf \langle i j\rangle }}
( {{ S^z}_{\bf i}}{{ S^z}_{\bf j}} - {1\over4} n_{\bf i} n_{\bf j} )
- {\rm t} \sum_{{\bf \langle i j \rangle},s}
({\bar c}^{\dagger}_{{\bf i},s} {\bar c}_{{\bf j},s} + h.c.) },
\tag 1
$$
\noindent where
${\rm  {\bar c}^{\dagger}_{{\bf i},s}}$
denote $hole$
operators;
${\rm n_{\bf i} = n_{{\bf i},\uparrow} + n_{{\bf i},\downarrow} }$
are number operators; ${\rm S^z_{\bf i}}$ are the z-components
of spins on the sites of the lattice;
and square clusters of ${\rm N}$ sites with
periodic boundary conditions are considered.
The rest of the notation is standard. If no holes are present, the
ground state is a N\'eel state with energy ${\rm E_{0h} = - J_z N}$.

To test the new algorithm, let us
first consider the case of one hole, i.e. ${\rm N - 1}$ electrons on an
${\rm N}$
sites cluster. As a preliminary exercise, the energies of clusters that
can be solved exactly using the Lanczos method,
namely ${\rm N \leq 26}$,\refto{didier,string,barnes}
were correctly  reproduced with
high accuracy. Following the
``truncation'' procedure described before, energies of larger clusters
can be obtained. In Fig.1a the energy
of the ground state in the one hole subspace is plotted as a function
of the size of the Hilbert space, for a cluster of ${\rm N=50}$ sites,
several coupling constants, and in the subspace of zero momentum.
Previous
studies of this model have shown that the ground state with one hole lies
in this subspace.\refto{string} The
starting state is a N\'eel state with one spin removed.
The 50 sites cluster corresponds to a tilted
square,\refto{oitmaa} and it is too large for a
conventional Lanczos study.
On Table 1, the energy as a function of the number of states in the
basis is presented for the most difficult case analyzed here,
i.e. ${\rm J_z = 0.2}$ and $50$ sites.
The convergence is clearly rapid, and
several digits of accuracy can be obtained without the need of
extrapolating results to the full Hilbert space.
For example, only $\sim 200,000$ states are enough to accurately
obtain the energies for the $6 \times 6$ cluster, while $\sim 600,000$ are
needed for the cluster of 50 sites to reach similar accuracy at $\jt = 0.2$.
This represents just a very small
fraction of the total basis set since, e.g.,  the cluster
of $50$ sites has a Hilbert space with $\sim 6 \times 10^{13}$ basis
states with definite momentum.
The best energies obtained with the new method are presented on Table 2
for the ${\rm N=36}$ and $50$ clusters, while the energies of smaller clusters
can be found in different places in the literature.\refto{string,didier}
Accurately knowing the ground state energies for finite clusters, an
analysis
as a function of the cluster size  provides the extrapolation to the
bulk limit.
Fig.1b shows the energies
as a function of the size of the cluster ${\rm N}$, for several coupling
constants.
The convergence with ${\rm N}$ is rapid and well described by an
exponential form,
$ E_N = E_{\infty} + A e^{-\alpha N} $ (for models with low energy
spin-waves, the convergence would be power-law in ${\rm N}$).
On Table 2, the ``bulk'' energy
is also shown. In all cases the error is in the last digit.

The fast convergence of the energy with the cluster size
observed in this model can be understood as follows:
Due to its mobility, a hole in an antiferromagnetic background creates a spin
distortion around it. This distortion has a typical radius, $r_s$, that
corresponds to the ``size'' of the wave function of the ground state,
and thus, in
average, the hole will be localized inside $r_s$ most of the
time.\refto{foot2} Then,
a good approximation to the total energy will be obtained
if the truncated Hilbert space is large enough to describe accurately
the movement
of the hole inside $r_s$, since
the background of spins at larger distances is frozen
into a N\'eel configuration on the ${\rm t-J_z}$ model. Including
basis states
where the hole has moved a distance much larger than $r_s$ from its original
position is
inefficient, since these excursions are exponentially suppressed.
The typical
size $r_s$ is a function of the coupling. Reducing the value of $\jt$,
$r_s$ increases  since the hole is more mobile.
This effect appears in Fig.1a since reducing $\jt$, a larger
Hilbert space is needed to achieve similar accuracy.
In the interval $0.2 \leq {\rm J_z/t} \leq 0.6$, and using the bulk
energies quoted on  Table 2, the energy of one hole, with respect to the zero
hole background, can be accurately fit as $e_{1h} = a + {\rm J_z} + b {\rm
J_z}^{\alpha}$ (t=1), where $a = -3.620$, $b=2.924$, $\alpha = 0.666$, and
the linear term comes from ${\rm -1/4 n_{\bf i} n_{\bf j} }$ in
the Hamiltonian.
This is in good agreement with the best previous fit available in the
literature which was obtained
from an analysis of clusters with only up to 26 sites.\refto{didier}
The exponent $0.66$ in the energy
is also consistent with the ``string'' picture widely discussed
in the literature of holes in ${\rm t-J}$-like models.\refto{string}
It is also worth noticing that Monte Carlo methods\refto{barnes}
have been applied to the study
of one hole in the ${\rm t-J_z}$ model, since for this particular case
there are no sign-problems. However, the statistical errors of the Monte
Carlo technique reduces the accuracy of its predictions. For example,
at $\jt = 0.2$ on a $6 \times 6$ cluster, the reported
Monte Carlo energy\refto{barnes}
is $( -2.431  \pm 0.006)t$ to be contrasted against the considerably more
accurate result of Table 2
for the same cluster, namely $-2.41951t$.

Consider now two holes in the ${\rm t-J_z}$ model. This case
is more
difficult than one hole. For example, Monte Carlo simulations
cannot avoid
the sign-problem. Standard Lanczos calculations are restricted to
lattices smaller than ${\rm N=26}$ sites, since the Hilbert space with two
holes is roughly ${\rm \sim N}$ times larger than with one hole.
However, in spite of these difficulties,
the new ``truncation''
technique produces highly accurate results. In Fig.2a, the
energy as a function of the size of the Hilbert space for the
50 sites cluster is shown, for
several values of the coupling.
The starting state corresponds
to a N\'eel state with two adjacent spins removed, in a zero momentum
plane wave.
The convergence rates are similar to
those observed for the case of one hole.
On Table 2, energies are presented, and on Table 3 a convergence example
is given.
The results for the $6 \times 6$ cluster were obtained
using $\sim 1,300,000$ basis states, while for the cluster of 50
sites, $\sim 2,200,000$ basis states were needed, both in the zero
momentum subspace. They represent only
a tiny fraction of the total Hilbert space of each cluster.\refto{foot1}
On Fig.2b, the size dependence of the two holes energies is shown as
a function of the cluster size. The convergence with ${\rm N}$ is rapid, and,
as in the case of one hole, an
exponential form fits accurately the data. This result can be understood
as follows: each hole forms a spin polaron as discussed before. Two
$independent$ or weakly interacting holes would be difficult to study,
since they will cover the whole lattice in their random walk motion, and
thus all states of the Hilbert space would be needed. However, it may
occur that the holes form a $bound$ state, and thus they become a single
bipolaron. Working in the center of mass of the bound state, this
hole ``molecule'' has a finite size, which is the analog of the $r_s$
size discussed before. Then, in the binding region, the technique may
provide results as accurate of those of one hole, and the convergence
with the lattice size should be exponential, as observed in practice
in Fig.2b.

Let us make this discussion more quantitative.
Using the energies for one and two holes, it is possible to calculate
the so-called ``binding energy'', defined as $\Delta_2 = e_{2h} -
2e_{1h}$, where $e_{nh} = E_{nh} - E_{0h}$, and $E_{nh}$ is the
ground state energy of $n$-holes. If $\Delta_2$ is negative, two
holes in the system have a tendency to form a bound state. If a finite
density of holes is considered, these pairs of holes may condense into
a superconducting state. Since $both$ the one and two holes energies
converge exponentially to the bulk limit, $\Delta_2$ can be accurately
evaluated in this model. $\Delta_2$ on individual clusters, as well
as $\Delta_2$ in the bulk limit obtained using the
 extrapolated energies for one and two
holes of Table 2, are shown in Fig.3 and Table 2. In the thermodynamic limit,
the binding energy is zero
at the ``critical'' coupling ${\rm J_z/t ]_c \sim 0.18}$,
below which the bound state disappears, and the numerical calculations
for two holes are more difficult.
Note that the results for $50$ sites, and those extrapolated
to the bulk limit are almost identical. The critical value can also
be obtained from the best fits in the bulk of the individual one and two holes
energies as a function of
the coupling. For two holes the best fit corresponds to
${\rm e_{2h} = -7.050 + 7.104 J^{0.728}_z}$ (t=1),
in the interval $0.2 \leq {\rm J_z/t} \leq 0.8$. For a bipolaron,
``strings'' of incorrectly flipped spins should influence
considerably
on the energy, and thus a ${\rm J_z^{2/3}}$ power-law would be expected as for
one hole.\refto{young}
However, the term ${\rm -1/4 n_{\bf i} n_{\bf j}}$, which cannot be
evaluated exactly for two holes, contributes approximately linearly in
${\rm J_z}$ and thus the power 0.728 is an effective exponent coming from the
combination of a linear term, and a string-like power-law.
Using all the information given above, it can be shown
that $\Delta_2 \sim
{\rm 0.19 - 2J_z + 7.104 J^{0.728}_z -5.848 J^{0.666}_z }$, which vanishes
at the critical value $0.183$, in good agreement with the previous
estimation.

The results have implications for the physics of the
${\rm t-J}$ model in two dimensions.
Lanczos calculations on small clusters\refto{young} have suggested
that binding of holes exists in this model only after a finite coupling
${\rm J/t}$ is reached.
Recently, Green Function
Monte Carlo (GFMC) simulations on large clusters for two
holes\refto{manousakis}
have been carried out at ${\rm J/t} = 0.4$, $0.7$, and $1.0$, and the results
extrapolated to
smaller couplings. A finite ${\rm J/t ]_c \sim 0.28}$ was reported
in the bulk limit.
The result of the present paper for the ${\rm t-J_z}$ model, i.e.
${\rm J_z/t ]_c \sim 0.18}$ is compatible with the GFMC
estimation. Actually, the simplest argument to suggest the presence of
binding in these models comes from counting the number of missing
antiferromagnetically aligned bonds in the ground state, when two holes
are placed
at short and large distances. A frozen N\'eel background
produces a larger binding energy than a fluctuating one, and thus
the critical coupling should be smaller for the
${\rm t-J_z}$ model than for the ${\rm t-J}$ model, in agreement with
the numerical results.
The existence of a finite ${\rm J_z/t ]_c}$
where binding starts, contributes to the discussion on phase
separation in these models. It has been argued that phase separation
exists in the ${\rm t-J}$ model for $all$ values of the
coupling.\refto{emery} However, Lanczos results,\refto{riera} Quantum Monte
Carlo simulations,\refto{adriana} as well as high temperature
expansions\refto{putikka} suggest that a finite coupling is needed to
phase separate. If the regime of phase separation is achieved by
forming clusters of holes of an increasing size, which varies
continuously
with the coupling, then ${\rm J_z/t ]_c}$ for two holes
binding, introduces a lower and $finite$ bound on the
coupling necessary to induce
phase separation in the ${\rm t-J_z}$ model. Since the inclusion of
spin quantum fluctuations can only $weaken$ the tendency to bind
and phase separate, the results of the present paper implies
that for the ${\rm t-J}$ model a $finite$ coupling is
also needed to induce phase separation.

Summarizing, in this paper an algorithm based on a systematic
reduction of the Hilbert space of a model has been presented. This
method of diagonalization in a reduced Hilbert space can be
considered as a systematic way of improving an initial trial
ground state, generating the ``optimal'' set of states for a
given basis set dimension. The
technique was applied to the two dimensional ${\rm t-J_z}$ model
where the energies of one and two holes on
large clusters of 50 sites where accurately calculated. That
allowed the evaluation of the binding energy. In the bulk limit,
it was found that
a critical coupling ${\rm J_z/t ]_c \sim 0.18}$ is needed
to induce two holes binding. Under mild assumptions, this
result implies that the ${\rm t-J}$ model in two dimensions phase
separates only after a $finite$ coupling is reached.

This work benefited from
useful conversations with A. Moreo.
J. R. is supported in part by
CONICET, Argentina, and the Antorchas Foundation. He also thanks the
Supercomputer Computations Research Institute (SCRI) for its support.
The computer calculations were done at the CRAY-2 of the
National Center for Supercomputing Applications, Urbana, Illinois.

\endpage

\references

\refis{foot3} For the case of one hole, the energies obtained from
Hilbert spaces generated by the straightforward application of the
hopping term are not significantly improved by eliminating the
states with small weight. The situation is different for two holes
where a better convergence is obtained if at each step the states with
less weight are discarded. In a preliminary study of the t-J model,
it was observed that this procedure leads to a considerable reduction
of the energy.

\refis{trugman} This procedure could be considered as a numerical
implementation on finite lattices of the method deviced by
S. Trugman, \prb 37, 1597, 1988; $ibid$, \prb 41, 892, 1990, to study
the Hubbard model in strong coupling with one and two holes. In this
approach the holes were allowed to move just a few lattice spacings, and
thus the resulting Hilbert space contained only a few hundred states.

\refis{foot2} Working in a basis of definite momentum is equivalent to
working in the center of mass frame, and thus in that situation the hole
appears as localized.

\refis{foot1} For the
special case of 50 sites, an extrapolation was made as a function of
the dimension of the Hilbert space, ${\rm H_D}$. The best fit was
obtained with a power-law dependence of the energy as a function of
${\rm H_D}$. The results quoted
on Table 2 are only $weakly$ dependent on the extrapolation method.

\refis{adriana} A. Moreo et al, \prb 43, 11442, 1991.

\refis{riera} J. Riera and A. P. Young, \prb 39, 9697, 1989.

\refis{young} E. Dagotto, J. Riera and A. P. Young, \prb 42, 2347, 1990.

\refis{oitmaa} The ``square'' lattices that can cover the bulk, and
also hold a perfect N\'eel order with periodic boundary conditions
are those satisfying ${\rm N = n^2 + m^2}$, where ${\rm N}$ is the
number of sites, and ${\rm n},{\rm m}$ are integers, with ${\rm n +m}$
even. ${\rm n = m = 5}$ corresponds to the largest lattice used in
this paper (see J. Oitmaa and D. Betts, Can. J. Phys. {\bf 56}, 897 (1978)).

\refis{review} E. Dagotto, Int. J. Mod. Phys. {\bf B 5}, 907 (1991).
E. Dagotto et al., \prb 45, 10741, 1992; and
references therein.

\refis{muller} J. Bednorz and K. M\"uller, Z. Phys. {\bf B 64}, 188
(1986);
A. Hebard et al., Nature (London) {bf 350}, 600 (1991), and references
therein.

\refis{emery} V. Emery, S. Kivelson and H. Q. Lin, \prl 64, 475, 1990.

\refis{putikka} W. Putikka, M. Luchini and T. M. Rice, \prl 68, 538, 1992.

\refis{wilson} {W. Wenzel and K. G. Wilson, \prl 69, 800, 1992.}

\refis{didier} D. Poilblanc, H. J. Schulz, and T. Ziman, \prb 46, 6435,
1992.

\refis{string} E. Dagotto, R. Joynt, A. Moreo, S. Bacci, and E.
Gagliano,
\prb 41, 9049, 1990.

\refis{barnes} T. Barnes et al., \prb 40, 10977, 1989.

\refis{manousakis} M. Boninsegni and E. Manousakis, FSU-SCRI preprint,
October 1992.

\refis{germans} H. De Raedt and W. von der Linden, \prb 45, 8787, 1992.

\endreferences

\endpage

\vskip 1cm

\bigskip
\centerline{\bf Table Captions}
\medskip

\item{1} Energy $E_{1h}$ of one hole in the ${\rm t-J_z}$ model, as a function
         of the size of the Hilbert space, ${\rm H_D}$, for a cluster of
         50 sites, and coupling $\jt = 0.2$.

\item{2} Energies of one and two holes in the ${\rm t-J_z}$ model, with
         respect to the energy of zero hole, for
         several couplings $\jt$. Quoted are the results for clusters
         of $6 \times 6$, and ${50}$ sites, as
         well as
         the extrapolated bulk energies obtained as described in the text.
         The last column denotes the binding energy, $\Delta_2$, in the
         bulk limit, as a function of ${\rm J_z/t}$.

\item{3} Energy $E_{2h}$ of two holes in the ${\rm t-J_z}$ model, as a function
         of the size of the Hilbert space, ${\rm H_D}$, for a cluster of
         50 sites, and coupling $\jt = 0.3$.

\vfill
\eject

Table 1
\vskip 0.4cm
\vbox{\offinterlineskip
\hrule
\halign{&\vrule#&\strut\quad\hfil#\quad\cr
\noalign{\hrule}
height2pt&\omit&&\omit&&\omit&\cr
&${\rm H_D}$\hfil&&$E_{1h}$\hfil&\cr
height2pt&\omit&&\omit&&\omit&\cr
\noalign{\hrule}
height2pt&\omit&&\omit&\cr
&49&&-12.263889&\cr
height2pt&\omit&&\omit&\cr
&141&&-12.354049&\cr
height2pt&\omit&&\omit&\cr
&1177&&-12.408446&\cr
height2pt&\omit&&\omit&\cr
&3389&&-12.414565&\cr
height2pt&\omit&&\omit&\cr
&9786&&-12.416875&\cr
height2pt&\omit&&\omit&\cr
&27484&&-12.417791&\cr
height2pt&\omit&&\omit&\cr
&77386&&-12.418091&\cr
height2pt&\omit&&\omit&\cr
&146587&&-12.418187&\cr
height2pt&\omit&&\omit&\cr
&247727&&-12.418213&\cr
height2pt&\omit&&\omit&\cr
&396726&&-12.418215&\cr
height2pt&\omit&&\omit&\cr
&666372&&-12.418222&\cr
height2pt&\omit&&\omit&\cr}
\hrule}

\vskip 0.7cm

Table 2
\vskip 0.4cm
\vbox{\offinterlineskip
\hrule
\halign{&\vrule#&\strut\quad\hfil#\quad\cr
&\hfil&&\multispan5\hfil 1 hole \hfil&&\multispan5\hfil 2 holes \hfil&\cr
\noalign{\hrule}
height2pt&\omit&&\omit&&\omit&&\omit&&\omit&&\omit&&\omit&\cr
&J_z/t\hfil&&$6 \times 6$\hfil&&$50$\hfil&&
$ bulk$\hfil &&$6 \times 6$\hfil&&$50$\hfil&&$bulk$\hfil&&$\Delta_2&\cr
height2pt&\omit&&\omit&&\omit&&\omit&&\omit&&\omit&&\omit&\cr
\noalign{\hrule}
height2pt&\omit&&\omit&&\omit&&\omit&&\omit&&\omit&&\omit&&\omit&\cr
&1.00&&0.28515&&0.28516&&0.28516&&0.0560&&0.0561&&0.0561&&-0.514&\cr
height2pt&\omit&&\omit&&\omit&&\omit&&\omit&&\omit&&\omit&&\omit&\cr
&0.80&&-0.30600&&-0.30598&&-0.30597&&-1.0105&&-1.0105&&-1.0106&&-0.399&\cr
height2pt&\omit&&\omit&&\omit&&\omit&&\omit&&\omit&&\omit&&\omit&\cr
&0.60&&-0.93918&&-0.93912&&-0.93910&&-2.1504&&-2.1510&&-2.1512&&-0.273&\cr
height2pt&\omit&&\omit&&\omit&&\omit&&\omit&&\omit&&\omit&&\omit&\cr
&0.40&&-1.63173&&-1.63151&&-1.63148&&-3.3965&&-3.3994&&-3.4012&&-0.138&\cr
height2pt&\omit&&\omit&&\omit&&\omit&&\omit&&\omit&&\omit&&\omit&\cr
&0.30&&-2.00974&&-2.00927&&-2.00924&&-4.0796&&-4.0861&&-4.0888&&-0.070&\cr
height2pt&\omit&&\omit&&\omit&&\omit&&\omit&&\omit&&\omit&&\omit&\cr
&0.25&&-2.20978&&-2.20903&&-2.20899&&-4.4430&&-4.4531&&-4.4570&&-0.039&\cr
height2pt&\omit&&\omit&&\omit&&\omit&&\omit&&\omit&&\omit&&\omit&\cr
&0.20&&-2.41951&&-2.41823&&-2.41821&&-4.8253&&-4.840&&-4.848&&-0.012&\cr
height2pt&\omit&&\omit&&\omit&&\omit&&\omit&&\omit&&\omit&&\omit&\cr}
\hrule}

\vskip 0.7cm

Table 3
\vskip 0.4cm
\vbox{\offinterlineskip
\hrule
\halign{&\vrule#&\strut\quad\hfil#\quad\cr
\noalign{\hrule}
height2pt&\omit&&\omit&&\omit&\cr
&${\rm H_D}$\hfil&&$E_{2h}$\hfil&\cr
height2pt&\omit&&\omit&&\omit&\cr
\noalign{\hrule}
height2pt&\omit&&\omit&\cr
&234&&-18.707940&\cr
height2pt&\omit&&\omit&\cr
&696&&-18.882805&\cr
height2pt&\omit&&\omit&\cr
&6204&&-19.026339&\cr
height2pt&\omit&&\omit&\cr
&18416&&-19.052528&\cr
height2pt&\omit&&\omit&\cr
&52672&&-19.066660&\cr
height2pt&\omit&&\omit&\cr
&106435&&-19.074957&\cr
height2pt&\omit&&\omit&\cr
&212486&&-19.079975&\cr
height2pt&\omit&&\omit&\cr
&673640&&-19.083531&\cr
height2pt&\omit&&\omit&\cr
&980681&&-19.084816&\cr
height2pt&\omit&&\omit&\cr
&1502829&&-19.085503&\cr
height2pt&\omit&&\omit&\cr
&2249454&&-19.085857&\cr
height2pt&\omit&&\omit&\cr}
\hrule}

\vfill
\eject

\vskip 1cm

\bigskip
\centerline{\bf Figure Captions}
\medskip

\item{1a} Ground state energy, $\Delta E_{1h}$, of the ${\rm t-J_z}$
          model measured with respect to the energy obtained in the first
          iteration of the truncation method, as a function
          of the size of the Hilbert space, ${\rm H_D}$. Here,
          a cluster of 50 sites and several ratios $\jt$ are considered.
          Note the rapid convergence of
          the results.
\item{1b} Ground state energy, $e_{1h} = E_{1h} - E_{0h}$,
          of the ${\rm t-J_z}$ model
          with one hole as a function of the number of sites of
          the cluster, ${\rm N}$, for several ratios, $\jt$. The energy is
          measured with respect to the zero hole ground state
          energy $E_{0h}$. The ``stars''
          denote the energies extrapolated to the bulk limit
          using the exponential form discussed in the text.

\item{2a} Same as Fig.1a but for the case of two holes.

\item{2b} Ground state energy, $e_{2h}= E_{2h} - E_{0h}$, of the ${\rm t-J_z}$
model
          with two holes as a function of the number of sites of
          the cluster, ${\rm N}$, for several ratios, $\jt$. The energy is
          measured with respect to the zero hole ground state
          energy $E_{0h}$. The stars
          denote the energies extrapolated to the bulk limit
          using the exponential form discussed in the text.

\item{3 } Binding energy, $\Delta_2 = e_{2h} - 2 e_{1h}$, as a function
          of $\jt$, for several cluster sizes. Starting from above, the
          open squares
          denote results for $N=16$, $20$, $26$, $36$, and $50$
          sites, respectively. The solid line is the binding energy
          extrapolated to the bulk limit, which is almost identical
          to the results on the $50$ sites cluster.

\endit